\documentclass[namedreferences]{SolarPhysics}
\usepackage{graphicx}        
\usepackage{color}           
\usepackage{url}             

\newcommand{\aap}{    {\it Astron. Astrophys.}}

\newcommand{\apj}{    {\it Astrophys. J.}}
\newcommand{\apjl}{   {\it Astrophys. J. Lett.}}

\newcommand{\jgr}{    {\it J. Geophys. Res.}}

\newcommand{\pasj}{   {\it Pub. Astron. Soc. Japan}}

\newcommand{\solphys}{{\it Solar Phys.}}

\begin{document}

\begin{article}

\begin{opening}

\title{Failed Eruption of a Filament as a Driver for Vertical \\ 
Oscillations of Coronal Loops} 


\author{T.~\surname{Mrozek}$^{1}$}
\institute{$^{1}$ Astronomical Institute, University of Wroc{\l}aw, 
				ul. Kopernika 11, 51-622 Wroc{\l}aw, Poland
				email: \url{mrozek@astro.uni.wroc.pl}}

\begin{abstract}
We present observations of a failed eruption of a magnetic flux rope recorded during the M6.2 flare of 14 July 2004. The observations were mainly made with TRACE 171 {\AA} and 1600 {\AA} filters. The flare was accompanied by a destabilization of a magnetic structure observed as a filament eruption. After an initial acceleration the eruption slowed down and finally was stopped by the overlying coronal loops. The observations suggest that the whole event is well described by the quadrupole model of a solar flare. The failed eruption stretched the loops lying above and then they were observed oscillating. We were able to observe clear vertical polarization of the oscillatory motion in TRACE images. The derived  parameters of the oscillatory motion are the initial amplitude of 9520 km, the period of 377 s, and the exponential damping time of 500 s. Differences between the existing models and the observations have been found. The analyzed event is the second sample for global vertical kink waves found besides the first by Wang and Solanki (2004). 
\end{abstract}


\end{opening}

\section{Introduction}
 

According to Gilbert {\it et al.} \cite{gilbert2007}, filament eruptions can be divided into three different types: full, partial, and failed. The eruption is full when over 90\% of plasma and magnetic structure is ejected and escapes from the Sun. The partial type is more complex and should be divided into two classes: A and B. Class A is the eruption of the entire magnetic structure with a small amount of mass. The reason of plasma deficit in the eruption is mass draining. Class B of partial eruption is observed when a part of magnetic structure erupts with a small amount of plasma. The failed eruption is the third type. It happens when the eruption of magnetic structure and mass is observed, but for some reasons the eruption stops and does not escape from the Sun. The possible mechanisms, other than solar gravity, that can stop the eruption are: forces within the erupting flux rope (Vr\v{s}nak, \citeyear{vrsnak1990}), reaching an upper equilibrium (Vr\v{s}nak, \citeyear{vrsnak2001}; Green {\it et al.}, \citeyear{green2002}), magnetic tension force and momentum exchange with the background plasma (Wang and Sheeley, \citeyear{wang1992};  Archontis and T\"or\"ok, \citeyear{archontis2008}), kink instability and stabilization of the erupting filament (Ji {\it et al.}, \citeyear{ji2003}; T\"or\"ok and Kliem, \citeyear{torok2005}).

The failed eruption was predicted by the so-called quadrupole model for the arcade flare (Uchida {\it et al.}, \citeyear{uchida1999};  Hirose {\it et al.}, \citeyear{hirose2001}). As noticed by Hirose {\it et al.} (\citeyear{hirose2001}): "{\it ...the upward motion of the dark filament pushed up in the expanding magnetic arcade may eventually be arrested by the overlying closed field}". In this model the overlying magnetic field is an essential part of the whole restructuring magnetic configuration which leads to a solar flare and a filament eruption. The role of the overlying magnetic field in the evolution of filament eruptions is also discussed in many theoretical papers investigating the initial phase of eruption (Amari  and Luciani, \citeyear{amari1999}; T\"or\"ok and Kliem, \citeyear{torok2005}; Fan and Gibson, \citeyear{fan2007}). The authors conclude that if the decrease of the coronal magnetic field with height is not steep enough then the eruption can be confined. 

Wang and Zhang (\citeyear{wang2007}) analyzed a group of X-class solar flares recorded by SOHO EIT (Delaboudini`ere {\it et al.}, \citeyear{delaboudiniere1995}) during the period of 1996--2004. The ratio of magnetic fluxes between low and high portions of the corona was calculated. The result obtained shows clearly that eruptive and confined events have significantly different values of the magnetic flux ratios, although the total overlying flux is similar. Namely, for confined events the ratio ${F_{\rm high}}/{F_{\rm low}}$ has larger values than for eruptive ones. Such a result strongly supports the scenario in which the eruption is stopped by the overlying magnetic structures. 

Liu (\citeyear{liu2008}) selected several eruptions and divided them into three groups: failed eruptions (FE), full eruption of kink instability (KI), and full eruption of torus instability (TI). The field overlying the eruption was calculated based on the potential-field source-surface (PFSS) model (Schatten {\it et al.},  \citeyear{schatten1969}; Altschuler and Newkirk, \citeyear{altschuler1969}; Hoeksema {\it et al.}, \citeyear{hoeksema1982}; Wang and Sheeley, \citeyear{wang1992}), and the decay index for the magnetic field was calculated. The result obtained shows that FE are connected with the values of the decay index smaller than KI and TI. An asymmetry in the magnetic field is another factor which is important in stopping the eruption. Liu {\it et al.}  (\citeyear{liu2009}) observed two failed eruptions which seemed to have occured  asymmetrically with regard to the overlying magnetic field. As the authors showed the magnetic force could be significantly larger far from the axis of an active region. Thus, the eruption can be stopped more easily in such a location. 

Recently, a new observational technique has become possible namely, simultaneous EUV observations made from three different angles with the use of SOHO and STEREO instruments. Shen {\it et al.} (\citeyear{shen2010}) observed five failed and one successful eruptions from the same active region with the use of this method. The authors confirmed the role of the overlying magnetic field in confining the eruption. Moreover, they found that the successful eruption was accompanied by the strongest flare. As all investigated eruptions were observed in the region with the same horizontal field strength, the authors concluded that the kinetic energy is an important factor which determines the confinement of the eruption. 

Eruptions and solar flares cause disturbances in coronal loops. After an eruption, disturbed coronal loops, while going back to the stationary position, may be observed oscillating. Various modes of oscillations are expected in loops and there exists a large base of such observations (Edwin and Roberts, \citeyear{edwin1983}; Nakariakov and Verwichte,  \citeyear{nakariakov2005}; Aschwanden, \citeyear{aschwanden2005}). 

Here we will focus only on the observations of global kink modes (Edwin and Roberts,  \citeyear{edwin1983}). Depending on the polarization of the oscillation plane, two types of global kink oscillations can be distinguished: vertical and horizontal. The first type of motion is observed when a loop oscillates in the plane of the loop. This causes the expansion and shrinkage of the loop, {\it i.e.}, the length of the loop changes. The second type of motion takes place when the loop oscillates in the plane perpendicular to the loop plane. In this case we observe the swaying and we do not observe any changes in its length.

There are any number of observations of horizontal global kink oscillations (Schrijver {\it et al.}, \citeyear{schrijver2002}; Aschwanden {\it et al.}, \citeyear{aschwanden2002}). It was found that only 6\% of the investigated flares have excited coronal loop oscillations and all oscillations were of horizontal type. The oscillatory motions of loops were observed during the time intervals from 7 to 90 min. The periods calculated are in the range of 2--33 min and the damping time was estimated to be 3--33 min. The oscillating loops had semi-lengths from 37 to 291 Mm. The amplitudes of the oscillations were observed in the range of 0.1--8.8 Mm. 

Up to the present there is only one observation of the global vertical oscillation of a coronal loop (Wang and Solanki, \citeyear{wang2004}). The loop was poorly visible in the images obtained with TRACE 195 {\AA} filter. However, it was possible to estimate the oscillation period (3.9 min), amplitude (7.9 Mm), and decay time (11.9 min). The correlation between the oscillatory motions and the brightness changes was also detected. Namely, the loop was brighter when it shrank, which can be understood in terms of compression; the vertical oscillations cause the change in the loop length. 

The observations of the M6.2 flare of 14 July 2004 presented here will reveal the second example of vertical oscillations of coronal loops. Moreover we were able to observe the initiation of oscillation caused by the eruption propagating through the solar corona.

\section{Observational Data}

The analyzed M6.2 flare of 14 July 2004 was observed in the active region NOAA 10646 near the west solar limb with heliographic coordinates N14W61. It was a relatively strong event of M6.2 GOES class. The maximum brightness in soft X-rays (SXR) was observed at 05:25 UT. There was no CME reported within several hours before and after the M6.2 flare, according to the CME list opened at http://cdaw.gsfc.nasa.gov/CME\_list.

The 30 cm Cassegrain telescope installed on board the {\it Transition Region and Coronal Explorer} (TRACE) (Handy {\it et al.}, \citeyear{handy1999}) gives the $1$ arcsec resolution ($0.5$ arcsec per pixel) with a $8.5$ arcmin $ \times 8.5$ arcmin field of view. The  wavelength selection is managed using the system of filter wheels giving possibility to observe the Sun in different bands from white light to EUV. Images are taken with the CCD coated with Lumogen to increase the efficiency of the detector for UV and EUV radiations  (Deeg and Ninkov, \citeyear{deeg1995}; Kristianpoller and Dutton,  \citeyear{kristianpoller1964}). 

The observational analysis of the M6.2 flare of 14 July 2004 was made mainly based on data from the TRACE telescope. TRACE obtained more than 90 EUV images with the 171 {\AA} filter during the time of the event. The temporal resolution of these observations varied from 8 s (impulsive phase and flare maximum) to 30--40 s (pre-flare activity and gradual phase). There are also UV images obtained with the 1600 {\AA} filter (temporal resolution 30--50 s). The UV images were collected during the pre-flare activity and decay phase. There were no UV data for the main phase of the flare. 

\begin{figure*}
	\includegraphics[width=1.0\textwidth]{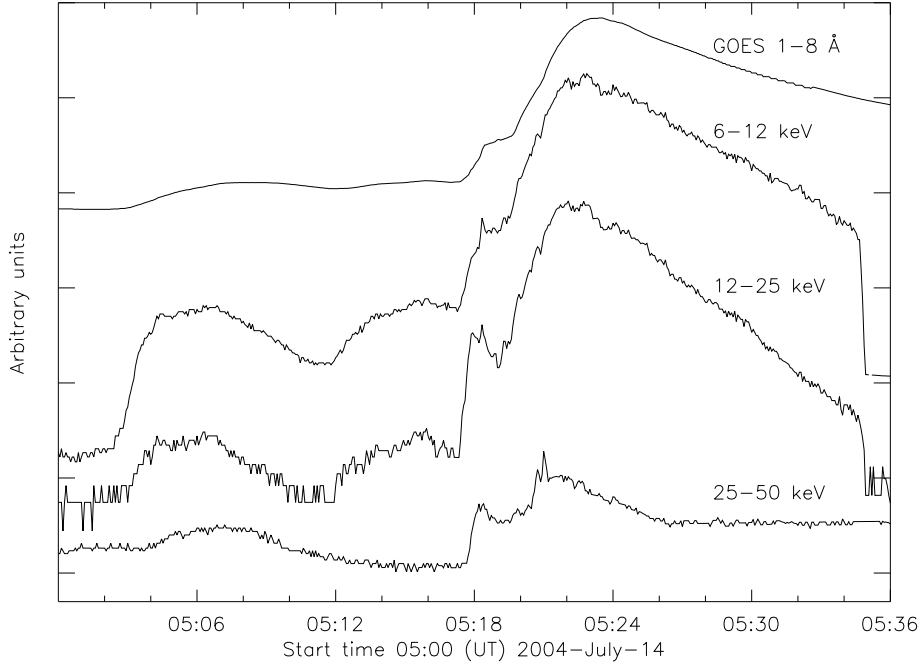}
  \caption{GOES soft X-ray flux profile and RHESSI X-ray light curves for the M6.2 flare of 14 July 2004. The curves are shifted vertically for better presentation.}
  \label{fig1}
\end{figure*} 

\begin{figure*}
	\resizebox{\hsize}{!}{\includegraphics{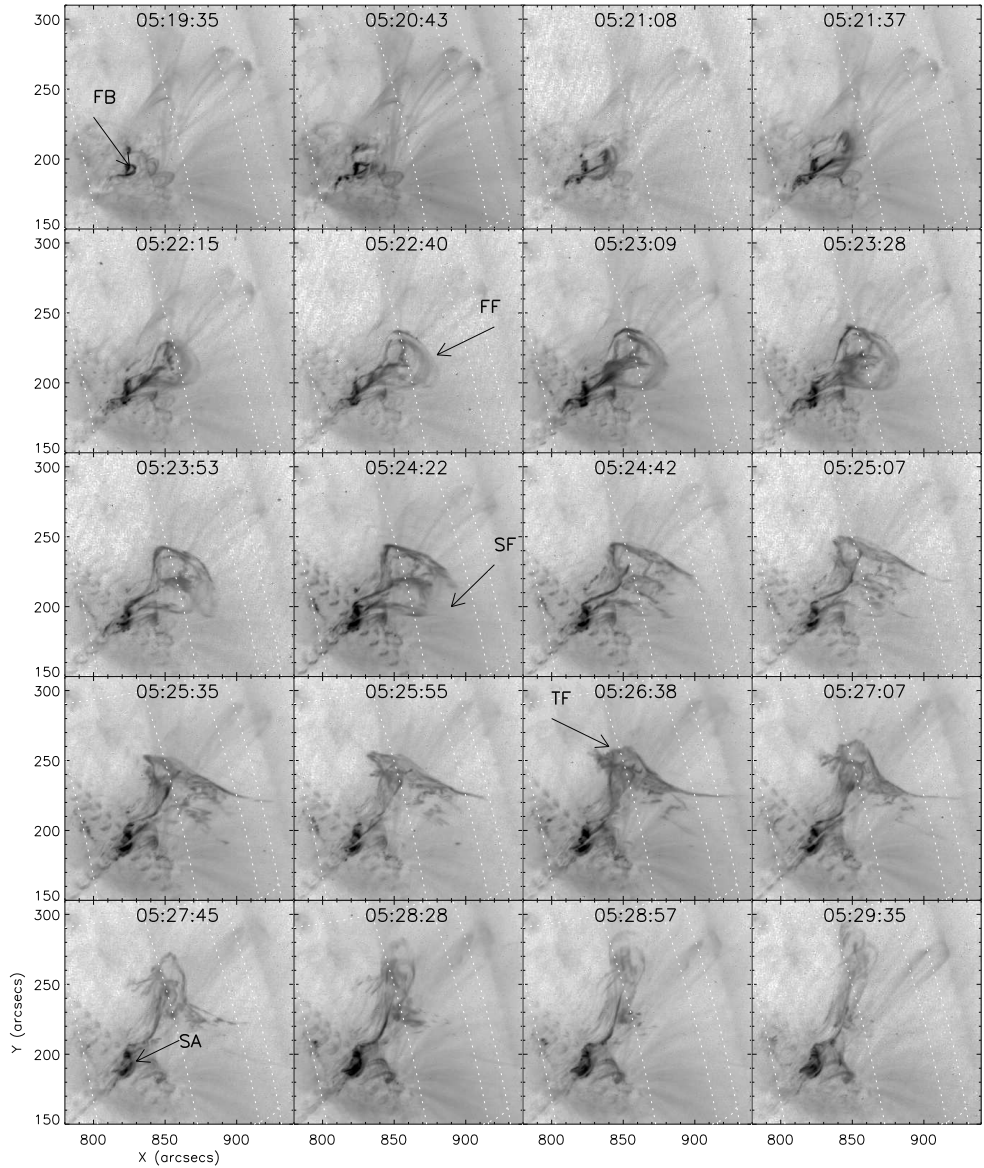}}  
	\caption{The evolution of the erupting structure. Images were taken with the TRACE 171 {\AA} filter. The first brightening (FB) was observed at 05:19:35 UT. The flaring structure was observed as an arcade of small loops (SA). Three distinct fronts can be observed within one erupting structure. The first front (FF) is visible starting from 05:21:37 UT. Its expansion was stopped at about 05:24:42 UT. At the same time the erupting structure was torn in the south part and the second front (SF) was observed. The third front (TF) was observed at 05:26:38 UT. It moved to the north and stopped at 05:30 UT. 
  }
	\label{fig2}
\end{figure*}

The temporal evolution of all the moving features was investigated with the use of the TRACE 171 {\AA} images. The projected height of the eruption was estimated through measuring the distance between the eruption front and the location of the first compact brightening visible in the top left panel of Figure~\ref{fig2}. During the whole evolution of the flare we observed a small arcade (SA) at the same place. There was a clear connection between this arcade and the erupting structure (Figure~\ref{fig2}). The height of the loops visible high in the corona was measured as the distance between the centroid of the loop-top and the baseline connecting the points where the loop was rooted. The baseline points can be determined by careful inspection of images obtained before and after the flare when the system of loops was brightest. 

The temporal evolution of the observed features was compared with the hard X-ray (HXR) light curves obtained by RHESSI (Lin {\it et al.}, \citeyear{lin2002}). Here we used the PIXON (Puetter and Yahil, \citeyear{puetter1999}) algorithm for HXR image reconstruction.

\section{Data Analysis}

The X-ray light curves from GOES and RHESSI instruments are shown in Figure~\ref{fig1}. Between 05:00 UT and 05:18 UT two small, gradual brightenings were visible. In this time interval, a group of small loops (whose height was about $10^4$ km) with increasing brightness was observed in EUV images. Moreover, strong brightening of footpoints was visible in HXRs, suggesting energy release in these small structures. The impulsive phase of the flare started at 05:18 UT and lasted for about 4 min. During this period the activation of a small magnetic rope followed by an  eruption was observed. During the rising phase we observed dramatic changes in the evolution of the analyzed flare, the eruption, and the overlying coronal loops. For clarity we divided the observational part of the analysis into two parts: evolution of the eruption and the vertical oscillation of the overlying loops.

\subsection{Evolution of the Eruption}

\begin{figure*}[!t]
	\resizebox{\hsize}{!}{\includegraphics{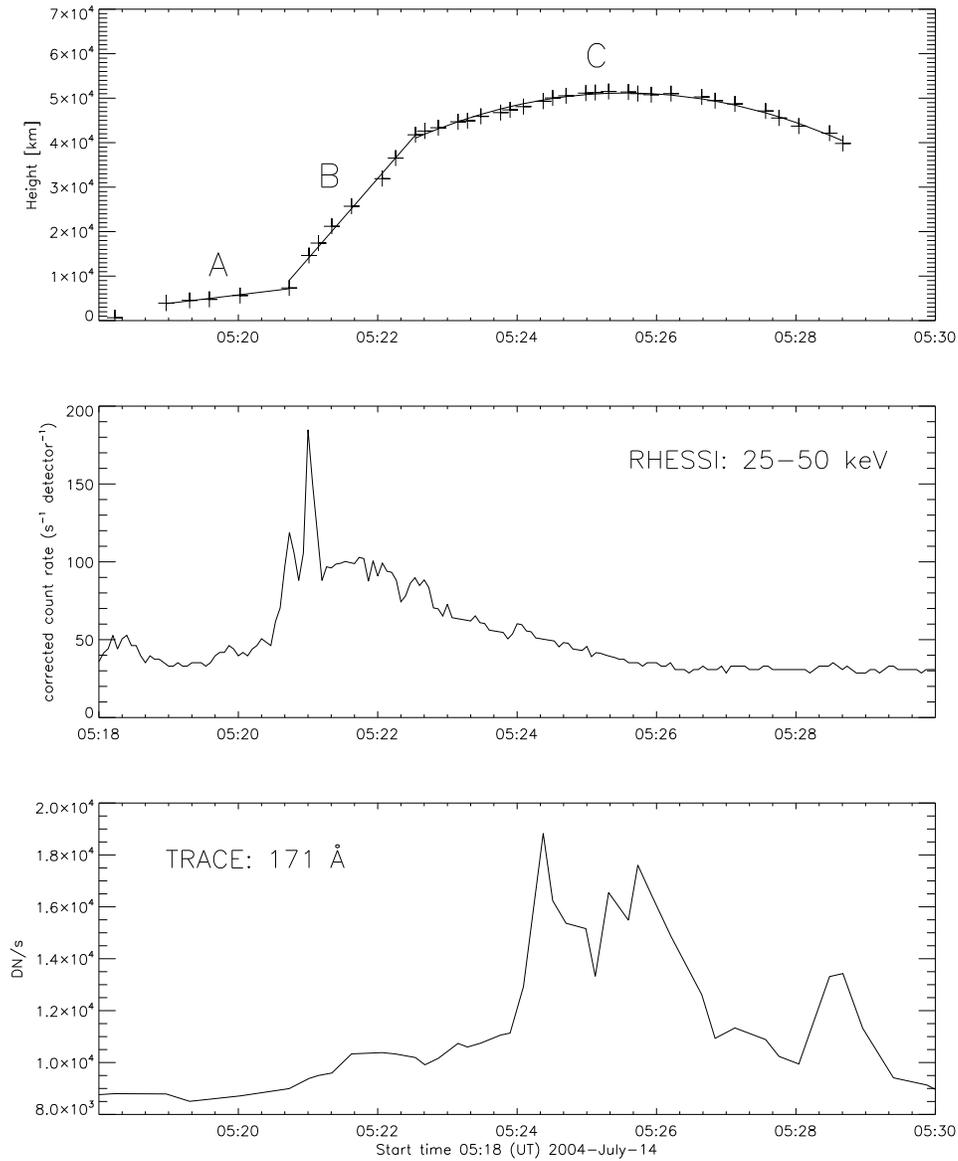}}
  \caption{Top panel: The evolution of the height of the eruption front with time. See text for the details of parameters of the fitted functions. Middle: the RHESSI light curve for  the 25-50 keV energy range. Bottom panel: the EUV light curve of one of the footpoint  observed during the impulsive phase. See text for more details.}
	\label{fig3}
\end{figure*}

During the first HXR burst (around 05:18 UT) we observed brightenings in the TRACE 
images and a small magnetic structure of increasing height (Figure~\ref{fig2}, top row). Next, during the strongest HXR peak (05:21 UT), a fast-moving structure was observed. Its characteristic blob-like shape can be seen in Figure~\ref{fig2} (second row). The shape changed dramatically two minutes after the eruption started. The front of the eruption got broken (Figure~\ref{fig2}, third row) and after that we observed a change in the direction of the propagation. Namely, at 05:24:42 UT, we observed the change of the direction to the south. Another change of the direction of the propagation took place at 05:26:38 UT. The structure was observed to move to the north. This third front was observed until it was stopped at about 05:30 UT. The changes of the projected height for the first front (FF) and the third front (TF) can be traced directly from the images (Figure~\ref{fig2}). The second front was very fast and faint, and we were not able to measure a reliable change of its height. The change of height for FF is presented in Figure~\ref{fig3} (top panel). Three significantly different stages are visible. 

The first stage (A in the top panel of Figure~\ref{fig3}) of the eruption evolution is connected with a very early phase of the flare when the eruption was observed as a small (3000 km in diameter) bright structure which slowly increased its height. The velocity of this initial rise was obtained from a linear fit to the change of height observed and was about $32\ \rm{km}\:\rm{s^{-1}}$. The value is similar to those obtained for the beginning phase from theoretical models of eruptions (T\"or\"ok and Kliem, \citeyear{torok2005}). 

The next stage, B, started at 05:20:40 UT. Its beginning is correlated with a strong HXR peak observed by RHESSI (Figure~\ref{fig3}, middle panel). The evolution of the height of B is also linear with a velocity of $300\ \rm{km}\:\rm{s^{-1}}$. This phase started very abruptly and we did not observe any acceleration phase between A and B. It looks rather like discontinuity in height changes. Moreover, the correlation with the HXR burst suggests that relatively strong magnetic field reconnection occurred and it caused a dramatic change in the velocity of the eruption propagation. Similar observations for the correlation between the acceleration phase of the CME and HXR peaks have been reported by Temmer {\it et al.} (\citeyear{temmer2008}).

A continuous deceleration of FF was observed since 05:22:40 UT until 05:29 UT when the foremost eruption front has been confined completely. From the quadratic fit to the observed changes of the projected height, we estimated the deceleration to be $620\ \rm{m}\:\rm{s^{-2}}$. This is about twice greater than the gravitational deceleration which means that some other mechanisms besides gravitation should be considered. 

\begin{figure*}[!t]
	\resizebox{\hsize}{!}{\includegraphics{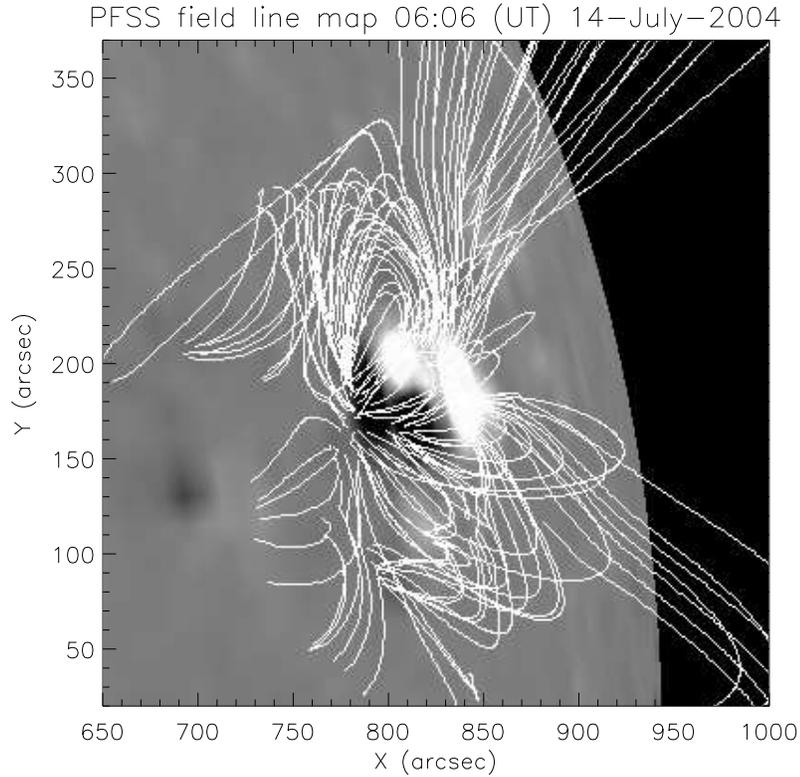}}
  \caption{PFSS reconstruction of the coronal potential magnetic field for the analyzed event. The background image presents the MDI data obtained about 40 min after the flare maximum.  }
	\label{fig4}
\end{figure*}

The observations suggest that the other system of loops was involved in braking the eruption. Namely, the event happened in the active region with clear quadrupole configuration. According to Hirose {\it et al.} (\citeyear{hirose2001}), such a configuration can develop an overlying field which effectively halts the eruption. Moreover, based on the PFSS model we have estimated the coronal magnetic structure in the analyzed region (Figure~\ref{fig4}). We used the standard PFSS package\footnote{http://www.lmsal.com/~derosa/pfsspack/} available under SolarSoftWare (SSW). Although it is a very simplified picture, a system of low-lying loops and high loops show a clear quadrupolar configuration. 

\begin{figure*}[!t]
	\resizebox{\hsize}{!}{\includegraphics{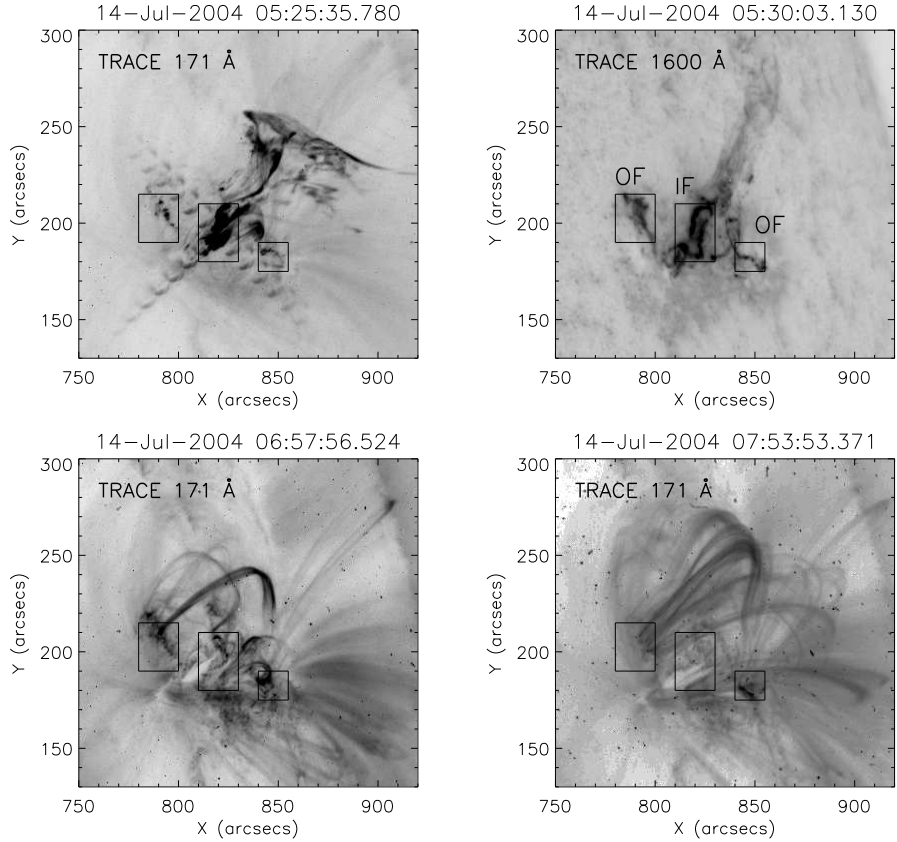}}
  \caption{Top left panel: EUV image of the analyzed event during the impulsive phase. Top right panel: UV image obtained immediately after the maximum of the flare. Two systems of footpoints (inner one, IF, and outer one, OF) were recognized and marked with boxes. See text for more details. Bottom panels: EUV images obtained one and two hours after the flare. The large system of loops was rooted in places which were recognized as OF during the maximum of the flare.}
	\label{fig5}
\end{figure*}

These overlying loops (OL) are visible in the TRACE EUV images two hours after the flare. The example images are presented in Figure~\ref{fig5} (bottom panels). The place where these loops are rooted to the chromosphere can be identified in these images. The footpoints can be recognized as bright areas in two top panels of Figure~\ref{fig5}. The top left panel shows the TRACE 171~{\AA} image obtained close to the flare maximum. An example of the TRACE 171~{\AA} image obtained during the decay phase is presented in the top right panel of Figure~\ref{fig5}. In the TRACE 1600~{\AA} image, two systems of footpoints are visible. The inner one (marked as IF in Figure~\ref{fig5}) correlates with the small arcade visible after the main phase of the flare. The outer ones (OF in  Figure~\ref{fig5}) are located in the footpoints of loops visible two hours after the flare maximum. Thus, we observe a clear quadrupolar configuration with a small arcade just after the reconnection and a large overlying system of loops. 

Additionally, we studied the changes of brightness of the selected areas. We chose the places where a large system of loops was rooted to the chromosphere. The analysis of brightness of the footpoints cannot be made in the TRACE 1600~{\AA} filter due to a lack of such observations in the main phase of the flare. However, there is no gap in the observation set taken with the TRACE 171~{\AA} filter. Both selected regions are marked with boxes in Figure~\ref{fig5} (top left panel) and designated as OF. The west footpoint is in the area that is greatly affected by the flare emission and by diffraction structures usually seen in the TRACE EUV images (Gburek {\it et al.}, \citeyear{gburek2006}). Fortunately, the east footpoint was located far from the inner footpoints (IF in the top left panel of Figure~\ref{fig5}) and between the arms of the diffraction cross. Thus, we were able to analyze the light curve of that footpoint. 

There was no significant change of brightness of the analyzed footpoint during the impulsive phase (Figure~\ref{fig3}, bottom panel). However, there were strong HXR bursts that should have been correlated with impulsive UV and EUV brightenings in the footpoints (Mrozek {\it et al.}, \citeyear{mrozek2007}), provided the connection between the acceleration region and the footpoint existed. Such strong brightenings were observed in the selected region between 05:24 UT and 05:29 UT only (Figure~\ref{fig3}, bottom panel). This is exactly the same time interval during which we observed the eruption slowing down (Figure~\ref{fig3}, top panel). It is possible that the brightenings were caused by non-thermal electrons which were produced in a higher system of loops during the interaction with the eruption. The main problem with such an interpretation is that we did  not observe the overlying loops during the flare. We can trace the changes of the footpoints, but the loops were visible two hours later (Figure~\ref{fig5}, bottom panels). However, the characteristic changes of the shape of the eruption suggest that some loops should have been present there. We did not find HXR sources in the place where the interaction between the eruption and the overlying field took place. The only HXR source visible in the images was correlated with the small flaring arcade.

\begin{figure*}
	\resizebox{\hsize}{!}{\includegraphics{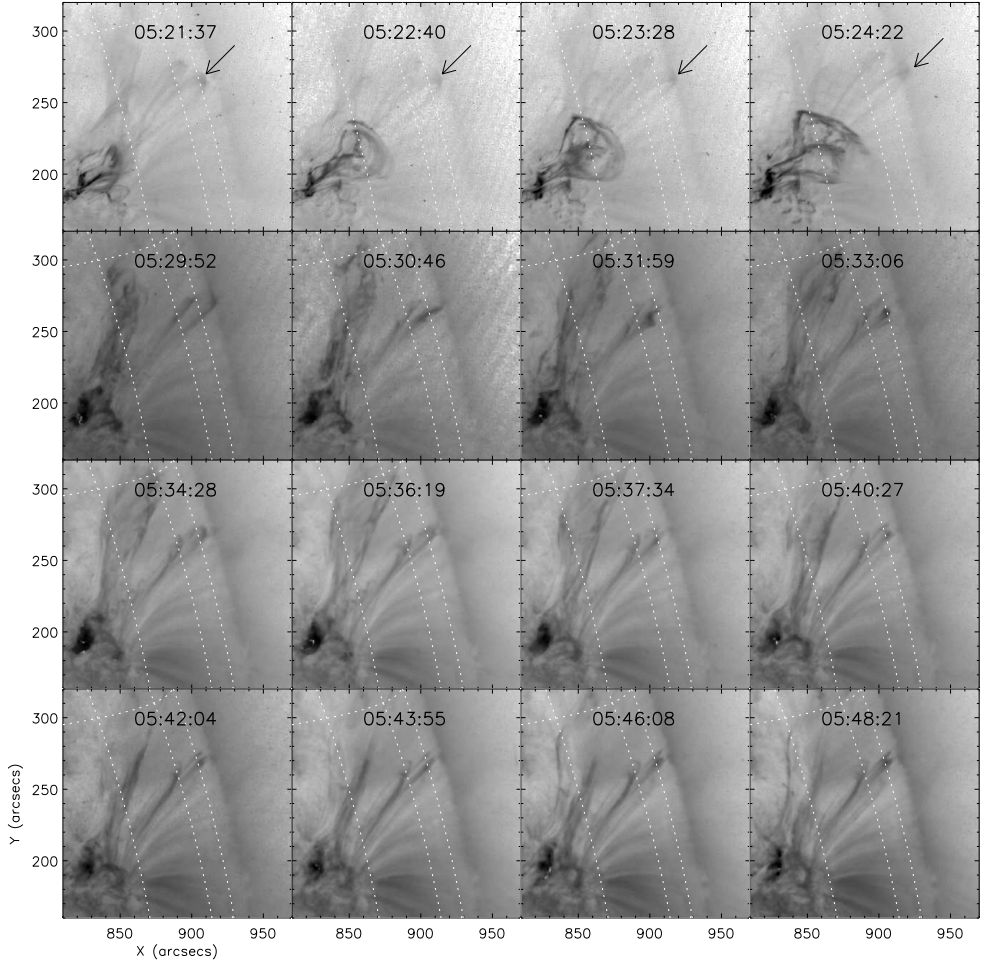}}  
	\caption{The changes in height of the second overlying system of coronal loops. The change of height for these loops is marked with an arrow in the top panels. In the next panels an oscillatory movement of these loops can be seen. }
	\label{fig6}
\end{figure*}

\subsection{Radial Oscillations}

The interaction between the eruption and the overlying field can be observed for even a higher system of loops (HL) which is marked with arrows in the top row of Figure~\ref{fig6}. We measured the height of this system from the pre-flare phase up to the late decay phase. Since the thickness of the loop top changed, we calculated the range of heights for the low and high boundaries of the loop-top source. This range of values is assumed to be a measure of uncertainty of the estimated heights. 

\begin{figure}
	\resizebox{\hsize}{!}{\includegraphics{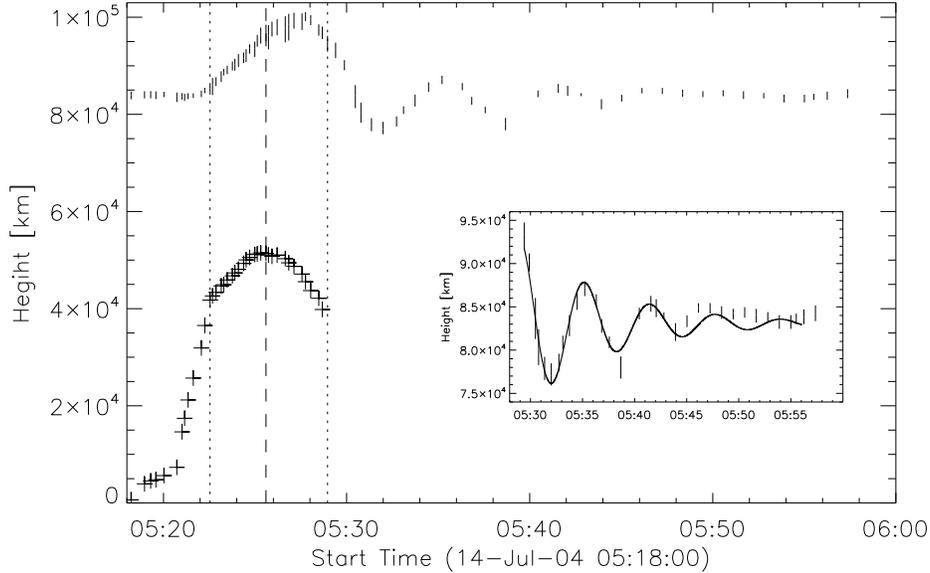}}
  \caption{The evolution of the high-lying loops is marked with vertical lines representing uncertainty in the location of the loop tops. The evolution of the   main eruption front is marked with pluses. The beginning and the end of deceleration of the eruption are shown with dotted lines. The dashed line marks the time for the maximum height of the eruption. The presented heights are absolute for both structures. Inset: The damped sine function (line) fitted to the measured height of the tops of high-lying loops (vertical lines). }
	\label{fig7}
\end{figure}

The change of height of HL is straightforwardly similar to the evolution of the eruption which can be seen in the top row of Figure~\ref{fig6}. Namely, during the development of the eruption the loops were pushed up also. The heights measured for the HL as well as the changes of height for the eruption front are shown in Figure~\ref{fig7}. All heights were obtained for the same reference height. We marked with dotted lines the start and end times of the deceleration of the eruption. The dashed line marks the time of the maximum height of the failed eruption. There is clear time shift between the maximum height of the failed eruption and the maximum height of the high-lying loops. This can be interpreted as an inertia effect of the loops pushed to greater heights. 

The eruption completely changed its direction after 05:29:50 UT (Figure~\ref{fig6}). Therefore, the force driving the movement of the high-lying loops disappeared, and the  loops started to fall free. They moved back and showed oscillations which are clearly visible in the TRACE images (Figure~\ref{fig6}). The oscillations are also visible in Figure~\ref{fig7} from 05:30 UT to 05:50 UT. The observed changes of height were fitted using a damped sine function: 

\begin{equation}
	H(t)=A\sin \left(\frac{2\pi}{P}t+\phi\right)e^{-\frac{t}{\tau}}+H_0
\end{equation}

where $A$ is the amplitude, $P$ is the period, $\tau$ is the characteristic time of damping. For the parameters $A$, $P$, and $\tau$ we obtained the values of 9520~km, 377~s, and 500~s, respectively. The fit is shown in the plot inserted in Figure~\ref{fig7}. It can be seen  that starting from 05:44 UT the fit is worse. It is possible that there were  actually more modes of oscillation excited in the observed loops that could explain this worse fit. However, we have to remember that during this very late phase the observed loops were weak and some additional errors in the estimated location might have arisen. 

The observed correlation between the changes of heights for the failed eruption and the HL shows that the latter was pushed up. Moreover, the observed event was located close to the solar limb which minimizes the projection effects. Thus, we expect that the observed disturbance of the loop tops was in the vertical direction, implying that the observed oscillations were also vertically polarized. There is one known observation of the global kink oscillations polarized vertically (Wang and Solanki, \citeyear{wang2004}). The authors obtained different parameters, {\it i.e.}, a lower value of period and a longer damping time with regard to the results shown here. As their oscillations wereobserved for one and a half period only, the obtained damping time was calculated with large uncertainty. 

Theoretical modeling (Selwa {\it et al.}, \citeyear{selwa2005}) shows that the $P/\tau$ ratio for global radial oscillations should be about 2.0. The observations of Wang and  Solanki (\citeyear{wang2004}) and the results reported here show that this ratio is significantly less than 1, implying that the models predict too large a damping of oscillations, {\it i.e.}, we did not observe as fast energy dissipation as the models predict. The second important difference between the models and the observations is the height of the oscillating loop before and after the oscillating motion. The models predict (Selwa {\it et al.}, \citeyear{selwa2005}) that after radial oscillations a loop has to be significantly higher. Wang and Solanki (\citeyear{wang2004}) reported that the loop had the same height before and after the oscillations. We observed the same effect (Figure~\ref{fig7}), {\it i.e.}, the height of the loops after the oscillations was exactly the same as before the oscillations ($8.4\times10^4$ km). This result suggests that there was no effect of 'stretching' the loop caused by oscillations. 

\section{Summary}

Here we showed the observation of the M6.2 flare of 14 July 2004. In this event two rarely observed features have met together, {\it i.e.}, failed eruption and radially oscillating loops. The investigated event was observed in a quadrupolar configuration of magnetic field. According to Hirose {\it et al.} (\citeyear{hirose2001}) in such a configuration we may expect that a developing eruption is stopped by the overlying system of loops. The  TRACE images and the PFSS model of coronal magnetic field show that we actually did observe two groups of coronal loops. One was a small arcade observed shortly after the flare. The second was a set of loops located higher and visible in the TRACE images one hour after the flare maximum. 

The interaction between the eruption and the above lying loops was observed in several ways:
\begin{itemize}
	\item The velocity and shape of the erupting material were observed to have changed. The change of the shape seemed to `trace' the change of the large system of loops observed after 06:00 UT.
	\item During the deceleration phase we detected impulsive EUV brightenings located at	the same position at which the OL were rooted.
	\item The deceleration of the eruption was significantly larger than the deceleration caused by  solar gravity.
	\item The height variations of the failed eruption and the HL have similar shapes.
	\item Side eruptions were observed in two directions where the OL were not present. It suggests that there exists a privileged region for an eruption to occur, correlated with the weak magnetic structures.
\end{itemize}

The observed failed eruption disturbed the HL observed at a height of $8\times10^4$ km. The loops were pushed upward to the height of $10^5$ km and then fall down and showed oscillations. The collected set of data allowed us to conclude that we observed vertically polarized oscillations. Global radial oscillations of loops were observed only once before (Wang and Solanki, \citeyear{wang2004}). The observed time changes of the height of the loops were fitted with a damped sine function. It allowed us to obtain the parameters of the oscillatory motion: namely, the initial amplitude of 9520 km, the period of 377 s, and the exponential damping time of 500 s. 

Significant differences between the existing models and the observations have been found. First, the ratio of the period to the damping time is not as large as the models predict (Selwa {\it et al.}, \citeyear{selwa2005}). Second, the oscillating loops have exactly the same height before and after the oscillations. There was no loop stretching due to oscillations as it is predicted theoretically.

The failed eruptions seem to be good events for causing disturbances in loops followed by radial oscillation of loops that are pushed to greater heights in the solar corona. In a typical erupting event the structure of the overlying loops is destroyed. Thus, this work will be performed for a larger group of events selected from the TRACE database. A detailed theoretical modeling should be also performed.

\section*{Acknowledgements}

The author is grateful to the TRACE and the RHESSI teams for providing great quality observational data. I acknowledge many useful and inspiring discussions of Professor Micha\l \mbox{ }Tomczak and also thank Barbara Cader-Sroka for editorial remarks. I am grateful to anonymous referee for a critical reading of the manuscript and useful comments. This investigation has been supported by a Polish Ministry of Science and High Education, grant No. N203 1937 33.



\end{article}

\end{document}